# Cache Optimization for Sharing Intensive Workloads on Multi-socket Multi-core servers

Murthy Durbhakula

Indian Institute of Technology Hyderabad, India

cs15resch11013@iith.ac.in, murthy.durbhakula@gmail.com

**Abstract.** Major chip manufacturers have all introduced multicore microprocessors. Multi-socket systems built from these processors are used for running various server applications. Depending on the application, remote cache-to-cache transfers can severely impact the performance of such workloads. This paper presents a cache optimization that can cut down remote cache-to-cache transfers. By keeping track of remote cache lines loaded from remote caches into last-level-cache and by biasing the cache replacement policy towards such remote cache lines we can reduce the number of cache misses. This in turn results in improvement of overall performance. I present the design details in this paper. I do a qualitative comparison of various solutions to the problem of performance impact of remote cache-to-cache transfers. This work can be extended by doing a quantitative evaluation.

## 1 Introduction

Many commercial server applications today run on cache coherent NUMA (ccNUMA) based multi-socket multi-core servers. Depending on the application, remote cache-to-cache transfers can impact overall performance. These are inherent to the application. One way to ameliorate this problem is to rewrite the application. Another way is to bias the cache replacement policy towards remote cache-to-cache transfers . In this paper, I present such a cache replacement policy for last-level caches (LLC) which will adaptively bias its replacement towards remote cache-to-cache transfers, when it's helpful for performance. We present a mechanism to observe the usefulness of bias and use it only when it's helpful. For instance, if there is a phase of application where there is little spatial/temporal locality of remote cache lines then we turn off the bias.  The rest of the paper is organized as follows: Section 2 presents the cache replacement policy. Section 3 briefly describes the qualitative methodology I used in evaluation. Section 4 presents results. Section 5 describes related work and Section 6 presents conclusions.

## 2 Cache replacement policy

In a ccNUMA system for every cache line there is a notion of home node. Home node is defined as the node in whose DRAM  a dirty cache line gets written back when it gets replaced. There are various ways home node gets determined for a given cache line. For instance, the upper bits of physical address of cache line determines the home node. Or it can be determined using a first-touch policy. In this paper I assume that the home node is determined by uppers bits of cache-line physical address. Further for cache-to-cache transfers I assume the hardware cache-coherence protocol to be MOESI (M-Modified, O-Owner, E-Exclusive, S-Shared, and I-Invalid).

In any last level cache in a given node/socket we maintain two remote-line-counters per cache set. One counter indicates the number of times we have biased the cache replacement policy for replacing remote cache lines whose owner is in another socket and home node is local node - **RemoteSharingLocalHomeCounter.**  Another counter indicates the number of times we have biased the cache replacement policy for remote cache lines whose owner is in another socket and home node is a remote node - **RemoteSharingRemoteHomeCounter.** We assume our last-level cache to be an inclusive cache where a cache line gets installed in LLC the very first time any of the cores request for the line and the line is not present in the local socket.

With every cache line in LLC we maintain a 1-bit storage which when set indicates that the line got installed in last-level-cache in "Shared" mode through a remote cache-to-cache transfer. This indicates that the owner of the line is a remote cache and the line is in "O" state in remote cache.

The default replacement policy for LLC is LRU. When replacing a cache line in LLC we see if the 1-bit sharing indicator of selected LRU line is set. If it is set then we see if the home node of the line is local or remote node by looking at the upper bits of physical address. If the home node is local node then we observe the value of **RemoteSharingLocalHomeCounter.** If the value of this counter is less than a specific threshold, say 1/4th of LLC associativity, then we replace another line which is not a shared line in LRU stack and increment the **RemoteSharingLocalHomeCounter.** On the other hand if the counter value is above the threshold then we replace the "shared" line and set the counter value to 0. If the home node is remote node then we observe the value of **RemoteSharingRemoteHomeCounter.** If the value of this counter is less than a specific threshold, say 1/2th of LLC associativity, then we replace another line which is not a shared line in LRU stack and increment the **RemoteSharingRemoteHomeCounter.** On the other hand if the counter value is above the threshold then we replace the "shared" line and set the counter value to 0. We have set different thresholds for **RemoteSharingLocalHomeCounter** and **RemoteSharingRemoteHomeCounter** because it is possible that the remote owner of the line could have done a silent write-back to main memory. In which case the penalty of accessing that line from local DRAM is less than accessing the same line from a remote DRAM. Further if the shared line is silently written back to main memory then biasing the cache replacement policy for such "shared" lines can potentially decrease performance. In the next sub-section I briefly present an adaptive mechanism and turn on and off the bias in cache replacement policy.

**Adaptive mechanism to turn on/off the cache replacement bias**

In a given time window T we also keep track of Remote_Miss_Fraction = (Number of Remote misses)/(Number of total misses). We use low and high water-mark for this metric. If this metric exceeds a high water mark then we turn on the cache bias. High water mark can be, for instance, 0.5 and low water mark can be 0.1. If it goes below low water-mark then we turn off the bias.

**3 Methodology**

I am using a qualitative methodology to compare the contributions of this paper with other existing approaches. Particularly I compared with hardware solutions and software solutions and the metrics I used are:

i) Ease of adaptation: That is, does the approach require changes from software or will it work seamlessly with existing software.

ii) Flexibility: That is, can the idea be improved or configured later on. Either software or hardware/software hybrid solutions have this advantage

iii) Verification complexity: Hardware solutions generally have verification complexity. They need to be fully verified before they can ship. Whereas software solutions can be potentially patched.

This work can be extended by doing a quantitative evaluation with various workloads.

## 4 Results

**Qualitative Comparison**

### 4.1 Hardware solutions:

**Remote cache in local DRAM**

Stanford FLASH project proposed using a portion of local DRAM as a cache for remote DRAM accesses, known as remote-access-cache (RAC) [1]. Similar to our solution this is a hardware solution that does not require changes in software. However, if the remote cache line working set is small then our approach of biasing cache replacement towards remote lines will let the remote line live in LLC longer hence resulting in lesser fetch latency than fetching the same line from local DRAM. Hence our solution works better for small to medium working set and local DRAM based remote cache works better for really large working sets.

### 4.2 Software solutions:

**Page replacement and migration**

OS based page replication and migration [2] has been proposed as a software solution for this problem. Page replication works for read-only data. And page migration works if the page is going to be modified by threads in only one socket. If page is shared in a producer-consumer manner across threads in different sockets and is going to be written by multiple threads of different sockets at different points of time then we cannot migrate the page to a specific socket.

**OS Scheduling optimization**

In my earlier work [3] I proposed optimizing OS scheduling algorithms to schedule threads such a way that the number of remote cache-to-cache transfers are minimized. Thus remote cache-to-cache transfers can become local cache-to-cache transfers. This works well if local cache-to-cache transfer latency is less than latency of LLC. Otherwise biasing LLC replacement works better.

| Solution | Need software changes | Flexibility | Hardware Verification Complexity |
|---|---|---|---|
| Biasing Cache Replacement | No | No | Yes |
| Remote Access Cache | No | No | Yes |
| OS based page migration and replication | Yes | Yes | No |
| OS Scheduling Optimization | Yes | Yes | No |

**Table 1: Comparison of Various Solutions**

**5 Related work**

I have already discussed some related work in the previous section. . In this section I am going to discuss some more. Natarajan and Chaudhuri [4] studied biasing of LLC cache replacement towards shared lines within multiple cores in a single socket. They show that cross-thread reuses of shared cache lines in LLC are more important than intra-thread reuses of private data in multi-threaded applications. And they also show that data sharing in LLC is greatly influenced by LLC replacement policy. Our study is complimentary to their study. Their focus is on single socket system whereas our focus is on multi-socket systems. We bias LLC replacement for shared remote lines which are owned by remote sockets. A miss to such a remote line would incur heavy penalty of fetching the line from a remote cache. And by biasing LLC replacement policy towards such remote lines we are saving on that penalty. Srikanthan et al [5] proposed a sharing-aware-mapper (SAM), a change to the operating system, that takes into account other factors such as latency tolerance and instruction level parallelism in addition to cache-to-cache transfers inherent in an application and then schedules threads accordingly. Whereas our solution is a pure hardware solution which works seamlessly with existing software. In cache-only-memory-architecture (COMA) [6] all of local DRAM is treated as a cache. This is also a hardware solution. However, for small to medium working sets our approach works better as we can serve remote data directly from cache instead of local DRAM. In my earlier work [7], I proposed biasing cache replacement policy of any one level of cache for remote DRAM lines which are not shared lines. Even there biasing last-level cache provides better tradeoff between size of working set and latency.

# 6 Conclusions

Many commercial server applications today run on cache coherent NUMA (ccNUMA) based multi-socket multi-core servers. Depending on the application, remote cache-to-cache transfers can impact overall performance. These are inherent to the application. In this paper I have presented a new LLC replacement policy based solution that can help reduce the performance impact of remote cache-to-cache transfers. I also presented an adaptive mechanism to observe the usefulness of bias and use it only when it's helpful. I have presented a qualitative evaluation of the solution. The solution presented in this paper does not require any software modifications. It works seamlessly with existing software.